\RequirePackage{lineno}
\documentclass[showpacs,aps,amsmath,amssymb,prc,showpacs,superscriptaddress,floatfix,preprint,preprintnumbers]{revtex4-1}
\usepackage{rotating}
\usepackage{epsfig}
\usepackage{hyperref}
\usepackage{lineno}

\usepackage{amsmath}
\usepackage{graphicx}
\usepackage[caption=false]{subfig}

\usepackage{xcolor}

\newcommand\snn{\ensuremath{\sqrt{s_{_{\rm{NN}}}}}}
\begin{document}

\title{Transverse momentum spectra of $f_0(980)$ from coalescence model}

\author{An Gu}
\email{gu180@purdue.edu}
\affiliation{Department of Physics and Astronomy, Purdue University, West Lafayette, IN 47907, USA}
\author{Fuqiang Wang}
\email{fqwang@purdue.edu}
\affiliation{Department of Physics and Astronomy, Purdue University, West Lafayette, IN 47907, USA}
\affiliation{School of Science, Huzhou University, Huzhou, Zhejiang 313000, China}
\date{\today}

\begin{abstract}
We use a coalescence model to generate $f_{0}$(980) particles for four configurations: ${s\bar{s}}$ meson, ${u\bar{u}s\bar{s}}$ tetraquark, $K\bar{K}$ molecule, and $u\bar{u}$ p-wave state. The phase-space information of the coalescing constituents is taken from a multi-phase transport (AMPT) simulation of proton-proton (pp) and proton-lead (pPb) collisions at the LHC. It is shown that the transverse momentum spectra and production yields of $f_0(980)$ differ significantly among the configurations. It is suggested that the $p_T$ spectra of the $f_0(980)$ compared to those of other hadrons (such as pion) and the ratio of the $f_0(980)$ $p_T$ spectra in pPb over pp collisions can be exploited to tell the configuration of the $f_0(980)$. 
\end{abstract}


\maketitle

\section{Introduction}

Exotic hadrons have configurations other than the usual ${q\bar{q}}$ and ${qqq(\bar{q}\bar{q}\bar{q})}$ configurations, and are a subject of wide interest. They are rare but allowed by quantum chromodynamics (QCD), their studies therefore hold great promise to improve our understanding of QCD \cite{Jaffe:2004ph}. The $f_{0}$(980), first observed in ${\pi\pi}$ scattering experiments in the 1970's~\cite{ Protopopescu:1973sh, Hyams:1973zf, Grayer:1974cr},  is one of the candidate exotic hadrons. Its configuration is not settled--- it can be a normal ${s\bar{s}}$ meson, a $u\bar{u}$ p-wave state, a $s\bar{s}g$ hybrid, a tetraquark ${s\bar{s}q\bar{q}}$ state, or a ${K\bar{K}}$ molecule~\cite{Bugg:2004xu, Klempt:2007cp, Pelaez:2015qba}.

Heavy ion collisions create a deconfined state of quarks and gluons, called the quark-gluon plasma (QGP) \cite{Adcox:2004mh,Arsene:2004fa, Back:2004je, Adams:2005dq, Muller:2012zq}. They provide a good environment to study exotic hadrons, because a large number of quarks and gluons permeate the QGP. When the temperature decreases, those quarks and gluons group into hadrons, presumably including exotic ones. This process is called hadronization and is not well understood. A common mechanism to describe hadronization in heavy-ion collisions is the quark coalescence in which several quarks(antiquarks) combine together to form a hadron \cite{Dover:1991zn,Fries:2008hs}. Coalescence model was originally developed to describe the formation of deuterons from targets exposed to proton beams \cite{Butler:1963pp} and is extensively used to describe hadron production in relativistic heavy ion collisions \cite{Dover:1991zn,Hwa:2002tu,Fries:2003vb,Greco:2003xt,Molnar:2003ff,Greco:2003mm,Fries:2003kq,Hwa:2004ng,Fries:2008hs,Minissale:2015zwa}.
It has been used to describe the NCQ scaling of elliptic flow, the baryon-to-meson ratio, and the hadron transverse momentum spectra, which can not be described well by fragmentation model~\cite{Fries:2003kq}. 

Previously we have used the coalescence model to study the elliptic anisotropy ($v_2$) of the $f_{0}$(980)~\cite{Gu:2019oyz}. We used the string melting version of the AMPT model (a multiphase transport)~\cite{Lin:2004en} model to produce the phase-space information of (anti-)quarks, as well as those of (anti-)kaons, and used those phase-space information as input to our coalescence. (AMPT uses quark coalescence to form hadrons \cite{Lin:2003jy,He:2017tla}, but it does not produce tetraquark or other exotic hadrons.) The $f_{0}$(980) is formed by coalescence with different configurations, and it was demonstrated that they yield distinctive $v_2$ following the number-of-constituent-quark (NCQ) scaling~\cite{Gu:2019oyz}. We argue that NCQ-scaling can be used to identify the $f_{0}$(980) quark content.

It has been shown that the $f_0(980)$ production yields and transverse momentum ($p_T$) spectra can also be used to probe the structure of the $f_0(980)$ \cite{Cho:2010db}.
%
In this work, we use the same coalescence model \cite{Gu:2019oyz} to calculate the $p_T$ spectra of the $f_{0}$(980). The parton phase space information is again generated by AMPT (with string melting) in pp and pPb collisions at nucleon-nucleon center-of-mass energy of $\snn=5.0$ TeV. 
The calculations are carried out for various quark configurations (${s\bar{s}}$, ${u\bar{u}s\bar{s}}$, ${K\bar{K}}$). 
We also check the $f_0(980)$ with $u\bar{u}$ p-wave state.
We could not check the $q\bar{q}g$ hybrid state as the AMPT model does not have gluons.
We demonstrate that the $f_{0}$(980) $p_T$ spectra can further help in identifying its quark content.

It should be noted, while quark coalescence is widely used as a hadronization model in heavy ion collisions, that its validity in small systems such as pp and pPb collisions is not without question. It has been shown in Ref.~\cite{Zhang:2021vvp}, however, that AMPT with proper parameters can describe particle spectra well in pp and pPb collisions. 
It is noteworthy that coalescence is usually considered applicable only at low to intermediate $p_T$, while at higher $p_T$ fragmentation must take over. In fact, a recent study~\cite{Zhang:2022fum} shows that independent fragmentation is needed in addition to coalescence in order to describe heavy flavor production in pPb collisions. 
In this exploratory work, we have strictly applied coalescence to pp and pPb collisions over the entire studied $p_T$ range. Our objective is to demonstrate the qualitative premise that the $p_T$ spectum of the $f_0(980)$ contains information about its quark content, but not aim to provide quantitative predictions of the spectrum.

\section{Coalescence Model}

The main idea of the coalescence model is to combine several partons into one hadron. 
Suppose N constituent particles are coalesced into a composite particle (a hadron or a ${K\bar{K}}$ molecule). The total yield of the composite particle can be expressed as \cite{Dover:1991zn}
\begin{eqnarray}
{N_{c}}=g_{c}\int \left( \prod_{i=1}^{N} \mathrm{d}N_{i} \right) f_{c}^{W}(\vec{r_{1}},\cdots, \vec{r_{N}},\vec{p_{1}},\cdots, \vec{p_{N}})\,, 
\label{equ:coalescence1}
\end{eqnarray}
where $f_{c}^{W}(\vec{r_{1}},\cdots, \vec{r_{N}},\vec{p_{1}},\cdots, \vec{p_{N}})$ is the Wigner function which is proportional to the coalescence probability,
and $g_{c}$ is the spin statistical factor. 



Partons closer in phase space are expected to have larger probabilities to coalesce into a hadron.
For two-body mesons, we take the Wigner function as Gaussian
\begin{eqnarray}
W_2(\vec{r_1},\vec{r_2}, \vec{p_1}, \vec{p_2})=g \cdot 8\exp\left(-\frac{r_{12}^2}{\sigma_r^2}-\frac{p_{12}^2}{\sigma_p^2}\right)\label{equ:mesona},
\end{eqnarray}
where
\begin{eqnarray}
r_{12}&=&|\vec{r_{1}}-\vec{r_{2}}|\label{equ:mesonb},\nonumber\\
p_{12}&=&\frac{m_1m_2}{m_1+m_2}\left|\frac{\vec{p_1}}{m_1}-\frac{\vec{p_2}}{m_2}\right|,\nonumber
\end{eqnarray}
and $g=1/4$ is the spin factor of a s-wave state.
Here, ${\vec{r_{i}}}$ and ${\vec{p_{i}}}$ are the position and momentum of $i$-th (anti-)quark at the time the hadron is formed. This calculation is implemented in the cluster center-of-mass frame. If we treat partons as Gaussian wave packets, the equivalent Wigner function would appear slightly different. The reader is referred to Ref.~\cite{Han:2016uhh} for details. Since partons in AMPT have determined position and momentum information simultaneously, we do not use the wave packet description. In AMPT, partons freeze out (FO, which means this (anti)quark doesn't interact with others anymore) at different times ${t_{i}}$. The moment for two or more (anti)quarks to coalesce is set to be the latest freeze out time of those (anti)quarks, ${t_{F}}$. The final positions are calculated as ${\vec{r}_{i}={\vec{r}}_{i,FO}+\vec{v}_{i,FO}\times(t_{F}-t_{i,FO})}$. For ${K}$ and ${\bar{K}}$ particles, we take their FO phase space information right after they are formed in AMPT. 

The spatial distance parameter ${\sigma_{r}}$ in Eq.~\ref{equ:mesona} reflects in general the size of the coalesced particle. It is given by $\sigma_{r}=1/\sqrt{\mu\omega}$, where $\mu=m_1m_2/(m_1+m_2)$ is the reduced mass of the two-body simple harmonic oscillator system, and $\omega$ is the oscillator frequency. The momentum distance parameter is given by $\sigma_{p}=1/\sigma_{r}$. 
The oscillator frequencies between different partons are from Ref.~\cite{Cho:2010db}. 
The quark masses are set to be the constituent quark masses ($m_u=m_d=0.3$ GeV, $m_s=0.5$ GeV). 
For $s\bar{s}$ quark system, the frequency is $\omega=0.519$ GeV and $\sigma_r=0.5477$ fm. 

For $K\bar{K}$ molecule, we can still use this Wigner function. The frequency is taken to be $\omega=0.0678$ GeV \cite{Cho:2010db} and the mass $m_K=0.4936$ GeV, so the parameter $\sigma_r=1.525$ fm. The $g$ factor is taken to be 1.

For a p-wave meson, the Wigner function is given by
\begin{eqnarray}
W_p(\vec{r_1},\vec{r_2}, \vec{p_1}, \vec{p_2})=g \cdot 8\left(-1+ \frac{2r_{12}^2}{3\sigma_r^2}+ \frac{2p_{12}^2}{3\sigma_p^2} \right)  \exp\left(-\frac{r_{12}^2}{\sigma_r^2}-\frac{p_{12}^2}{\sigma_p^2} \right) \label{equ:uubar}.
\end{eqnarray}
The spin factor is $g=3/4$.
For $u\bar{u}$ p-wave system, the frequency is $\omega=0.550$ GeV \cite{Cho:2010db}, thus $\sigma_r=0.6870$ fm.

For multi-particle systems, the quantum state is difficult to compute analytically. For tetraquark and pentaquark hadrons, only the heavy quark sector has been calculated quantum mechanically using perturbative approaches \cite{Ali:2017jda,Chen:2016qju,PhysRevD.57.6778,Shi:2013rga}. A widely used way to calculate the Wigner function can be found in \cite{Han:2016uhh}, which is based on the assumption, in the case of a baryon as an example, that the two partons form a simple harmonic oscillator (SHO) first, and then another SHO is formed with the third parton. In this work, for tetraquark systems, we treat all constituent quarks equally pair-wide and define the Wigner function to be Gaussian: 
\begin{eqnarray}\label{eq3a}
{W_4(\vec{r_{1}},\vec{r_{2}},\vec{r_{3}},\vec{r_{4}},\vec{p_{1}},\vec{p_{2}},\vec{p_{3}},\vec{p_{4}})}= g \cdot 8^3 \exp \left( -\frac{1}{ \sigma_{r}^{2}}{\sum\limits_{i,j=1 \ i < j}^{4} {r}_{ij}^{2}}-\frac{1}{ \sigma_{p}^{2}} {\sum\limits_{i,j=1 \ i < j}^{4} {p}_{ij}^{2}} \right). 
\label{equ:mesoncc}
\end{eqnarray}
For the tetraquark $s\bar{s}u\bar{u}$ component, we assume the parameter $\sigma_r$ to be the same as $s\bar{s}$ system, i.e.~$\sigma_r=0.5477$~fm. The spin factor is $g=1/8$.

Table~\ref{tab:coal} summarizes the Wigner Function coalescence parameters used in our study.
\begin{table}
\begin{center}
\caption{\label{tab:coal}Wigner Function coalescence parameters used in our study. The constituent quark masses ($m_u=m_d=0.3$~GeV and $m_s=0.5$~GeV) are used in the coalescence model.}
\begin{tabular}{c c c c c}
\hline
\rule{0pt}{10pt} System & Reduced mass $\mu$ (GeV) & $\omega$ (GeV) & $\sigma_r=1/\sqrt{\mu \omega}$ (fm) & $g$ factor \\
\hline
\rule{0pt}{10pt}
$u\bar{d}$          & 0.15      & 0.550     & 0.6870    & 1/4 \\
$u\bar{u}$          & 0.15      & 0.550     & 0.6870    & 3/4\\
$u\bar{s}$          & 0.1875    & 0.519     & 0.6324    & 1/4\\
$s\bar{s}$          & 0.25      & 0.519     & 0.5477    & 1/4\\
$u\bar{u}s\bar{s}$  & 0.25      & 0.519     & 0.5477    & 1/8\\
$K\bar{K}$          & 0.2468    & 0.0678    & 1.525     & 1\\
\hline 
\end{tabular}
\end{center}
\end{table}

In coalescence, the three momentum of the hadron is calculated as the sum of the three momenta of its constituents. The physical mass of the hadron is however directly assigned. The energy is not conserved in this process. It is pointed out in Ref.~\cite{Butler:1963pp} that a third body may cure this deficiency. It was also pointed out that the uncertainty principle may be considered to solve this problem \cite{Mrowczynski:1987oid}. 

\section{Results}

We generate approximately $5\times10^6$ minimum-bias (MB) events each for pp and pPb collisions at $\snn=5.0$~TeV. We take the (anti)quark phase space information from AMPT at freezeout after the parton cascade as input to the coalescence model. 
We carry out the coalescence calculation separately for each of the possible configurations of the $f_0(980)$. 
For each configuration, if the (anti)quarks flavors match the configuration and the value of a generated random number (uniformly distributed between 0 and 1) is smaller than the value of the Wigner function, the $f_0(980)$ is formed. 
Those (anti)quarks are then removed from further consideration in the coalescence calculation.

\begin{figure}
\subfloat{\includegraphics[width=0.5 \linewidth]{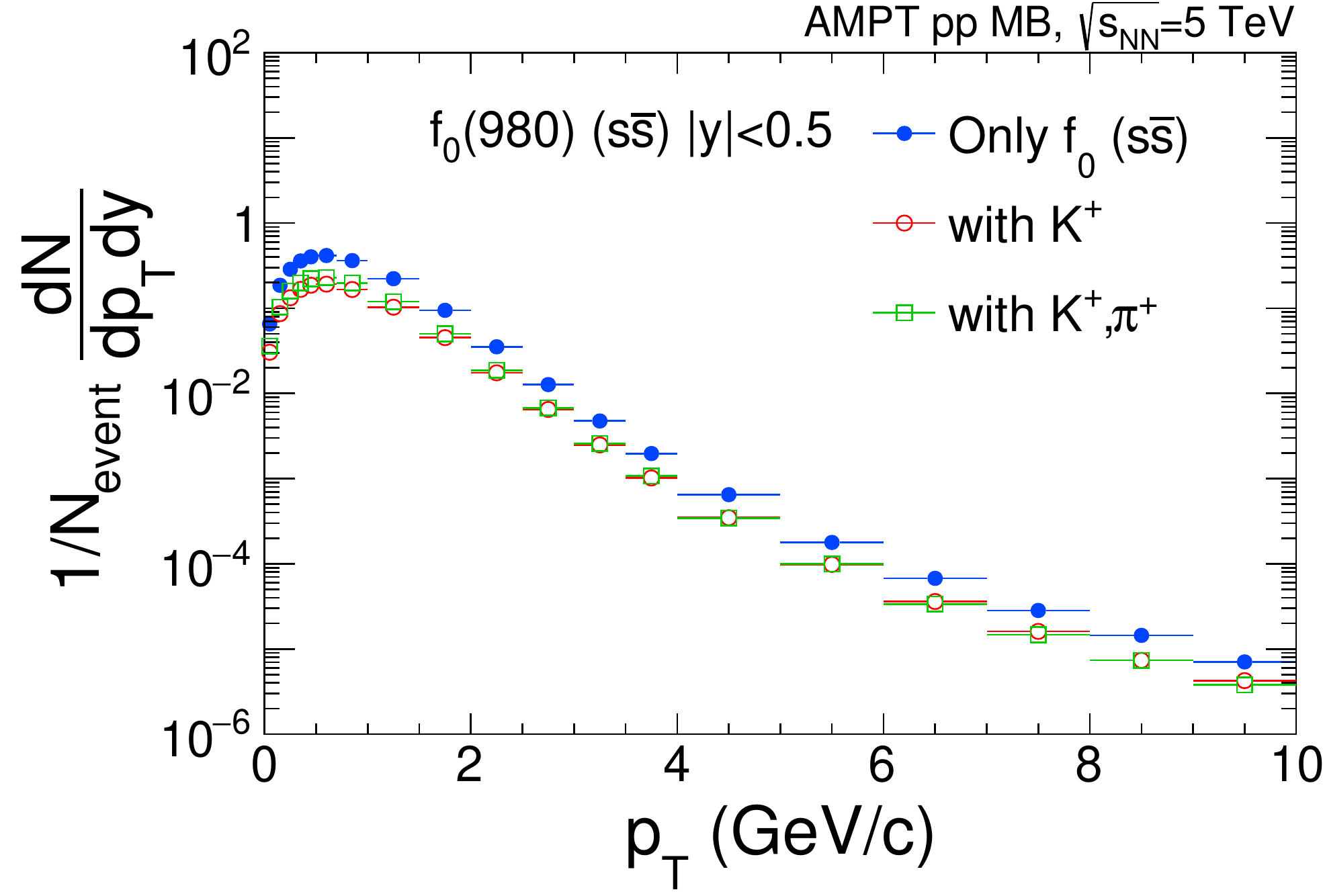}}
\subfloat{\includegraphics[width=0.5 \linewidth]{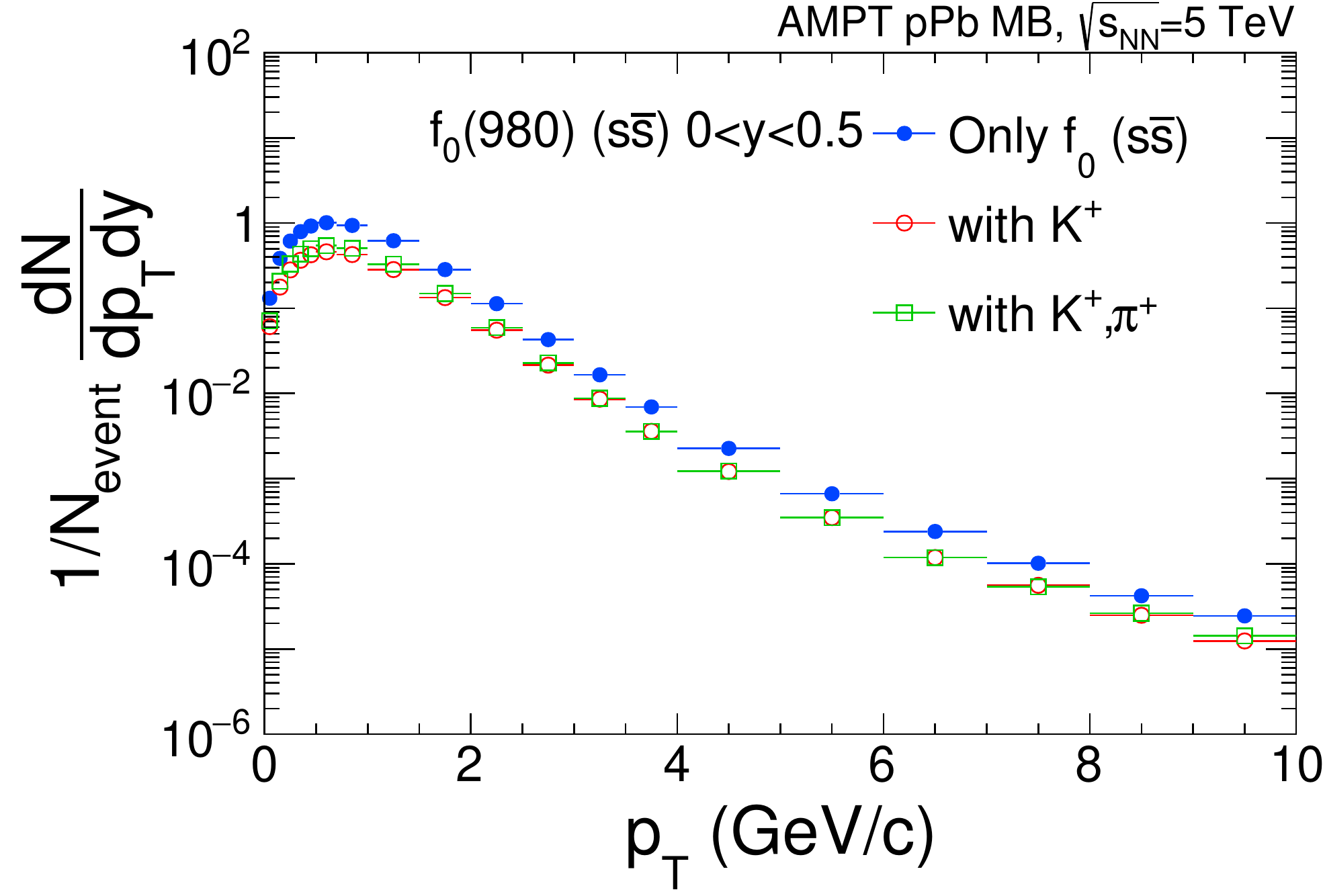}}
\caption{\label{fig:pt_compk}The $p_T$ spectrum of the $f_0(980)$ with $s\bar{s}$ configuration calculated by coalescence including formation of $K^+$ and $\pi^+$. The left panel is the result for pp MB events and right for pPb MB events. The input (anti-)quark phase-space information is taken from the corresponding AMPT simulations.}
\end{figure}

In reality, not only $f_0(980)$ can be formed but also other hadrons. 
For example, the formation of $K^+$ ($u\bar{s}$) will take some $\bar{s}$ quarks and reduce the yield of $f_0(980)$.
To study this effect, we check the yield of $f_0(980)$ with $s\bar{s}$ assumption while including the formation of $K^+$. 
For each $\bar{s}$ quark, we check whether there is a $u$ quark nearby in phase-space to form a $K^+$ by calculating the Wigner function. 
The $\omega$ parameter value for the $K^+$ is taken to be as same as that of the $s\bar{s}$ system, i.e. $\omega_s=0.519$ (GeV) \cite{Cho:2010db}. The reduced mass for the $u\bar{s}$ system ($K^+$) is $0.1875$~GeV, thus the $\sigma_r$ value is 0.6324 fm.
As usual, if a $K^+$ is formed, then the $\bar{s}$ and $u$ quarks are removed from the sample. The results are shown in Fig.1 as the red points.
For comparison, the blue points shows the $p_T$ spectrum of the $f_0(980)$ ($s\bar{s}$) without considering formation of other hadrons. 
Including the formation of $K^+$ ($u\bar{s}$) reduces the yield of $f_0(980)$ by about $30\%$. 
Since $u$ quarks can also be taken by other hadrons, e.g. $\pi^+$ ($u\bar{d}$), the yield of $K^+$ is likely smaller, so taking away fewer $\bar{s}$ quarks. 
This would increase the formation of $f_0(980)$. 
We thus check the $f_0(980)$ yield by including both $\pi^+$ ($u\bar{d}$) and $K^+$ ($u\bar{s}$) in coalescence. The results are shown in Fig. \ref{fig:pt_compk} as green points. 
The resultant yield is about $10\%$ higher than that obtained by including only $K^+$ in addition to the $f_0(980)$ in the coalescence, and consequently $20\%$ lower than the only $f_0(980)$ case (blue). 
We have checked the sequence of hadrons to form in coalescence and found negligible effect. 
Similarly, we have also checked the effect of looping over AMPT partons in its output order, reversed order, or randomly, and found negligible effect. 
Since the effect of including other hadrons in the coalescence is small compared to the $p_T$ spectrum difference among configurations and likely affects the various configurations and different collision systems in similar ways, we present results coalescing only the $f_0(980)$.

\begin{figure}
\subfloat{\includegraphics[width=0.5 \linewidth]{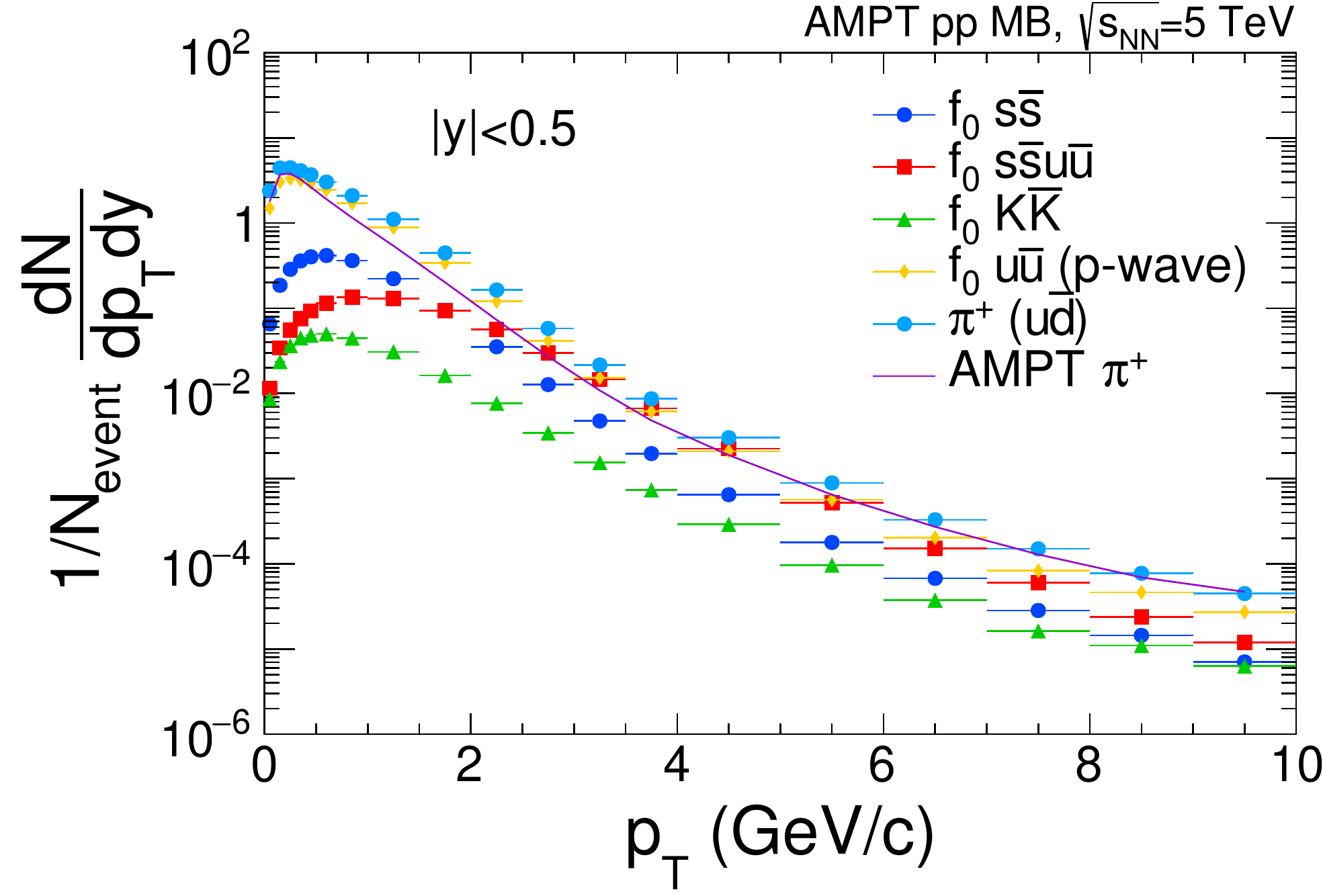}}
\subfloat{\includegraphics[width=0.5 \linewidth]{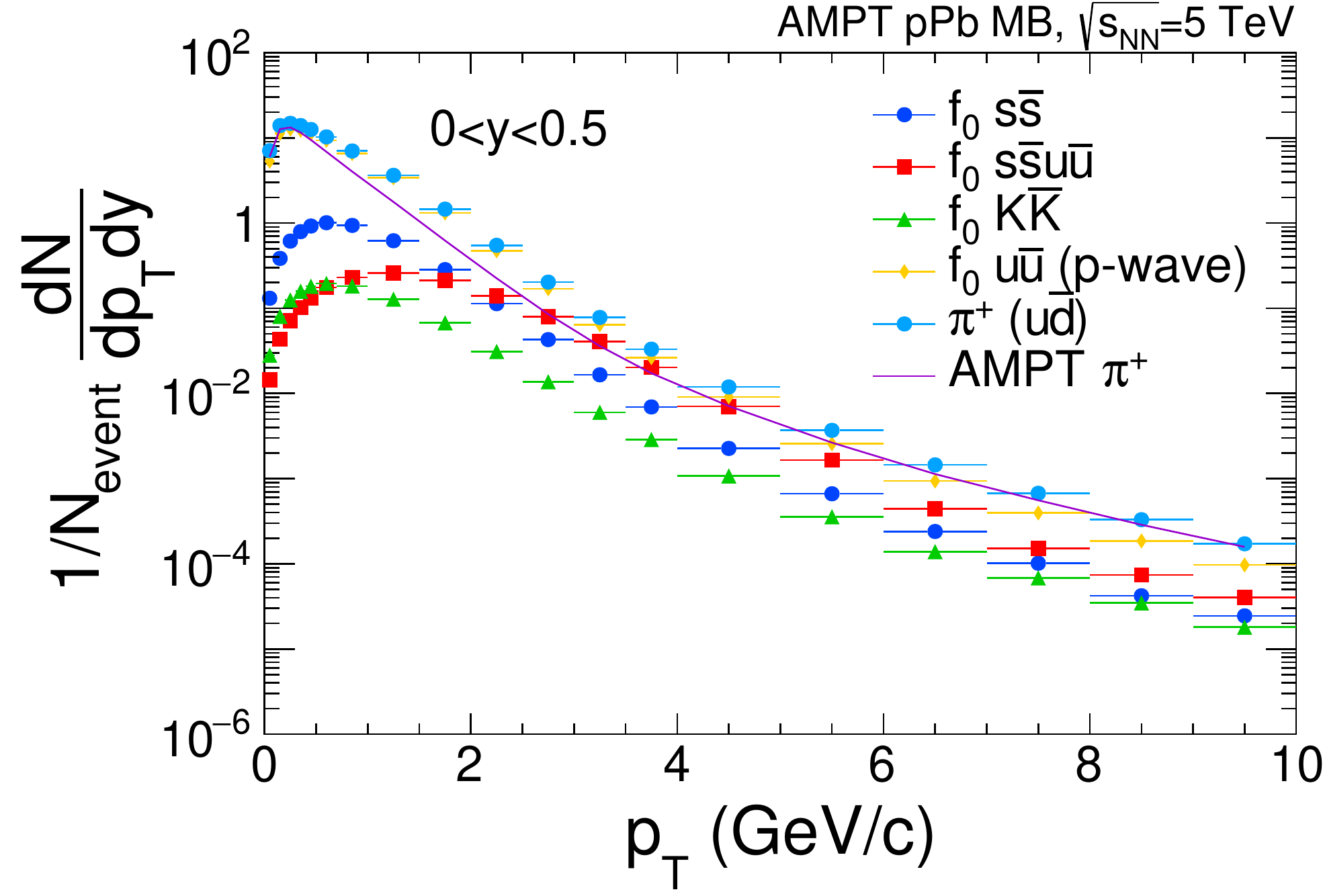}}
\caption{\label{fig:pt}The $p_T$ spectra of the $f_0(980)$ with different configurations. For comparison, the $\pi^+$ spectra from coalescence and directly from AMPT (curve) are also shown. The left panel is the result for pp MB events and right for pPb MB events. The input (anti-)quark phase-space information is taken from the corresponding AMPT simulations.}
\end{figure}

We use the coalescence model to generate the $f_0(980)$ of four different configurations ($s\bar{s}$, $u\bar{u}s\bar{s}$, $K\bar{K}$, $u\bar{u}$ p-wave). 
Figure.~\ref{fig:pt} show the $p_T$ spectra of the $f_0(980)$ of $s\bar{s}$, $u\bar{u}s\bar{s}$, $K\bar{K}$, $u\bar{u}$ (p-wave) configurations for pp and pPb collisions. 
As a sanity check, we also calculate the $\pi^+$ spectrum with our coalescence model and compare it to the $\pi^+$ spectrum directly from AMPT. 
This is shown in Fig.~\ref{fig:pt} as light blue points and purple curve, respectively. 
They are understandably not identical because of the different ways of implementing coalescence and because AMPT has the full collision evolution. 
However, the overall shapes and magnitudes are similar, lending confidence to our coalescence calculation. 
We have also compared the calculated $\pi^+$ to the ALICE measurement~\cite{ALICE:2019hno} and found reasonable agreement. 
For reference, the total $\pi^+$ yield from our coalescence calculation is 4.05, compared to 2.68 from AMPT and 2.07 from ALICE data~\cite{ALICE:2019hno}, with negligible statistical uncertainties. 
The consistency in the yields within a factor of $\sim2$ is reasonably good given our unsophisticated coalescence model and the question of validity to describe particle production in pp collisions entirely by coalescence. 

As shown in Fig.~\ref{fig:pt} the yield of $u\bar{u}$ p-wave states is the highest because there are more u (and $\bar{u}$) quarks than s (and $\bar{s}$) quarks. Its $p_T$ spectrum looks similar to that of AMPT $\pi^+$, while the yield is slightly lower because of its p-wave configuration. 
The $s\bar{s}$ state has a harder spectrum at low $p_T$ than the $u\bar{u}$ p-wave state, whereas their shapes are similar at higher $p_T$.
The $p_T$ spectrum of the $K\bar{K}$ molecule state appears to be harder than the $s\bar{s}$ state with a lower yield. This is understandable because the $K\bar{K}$ configuration has two more light quarks rendering its relatively rarer probability and larger $p_T$.
It is interesting to note that the tetraquark state spectrum lie between those of the $s\bar{s}$ and $K\bar{K}$ states at low $p_T$, and takes over at higher $p_T$. The comparison to the $s\bar{s}$ state may be relatively easy to understand: the probability to form a tetraquark state is significantly smaller so its yield, dominant at low $p_T$, is smaller, and the two extra light quarks make the tetraquark state significantly harder overtaking the $s\bar{s}$ state at high $p_T$.
It is intriguing to find that the tetraquark state is higher than the $K\bar{K}$ state except at very low $p_T$.
This arises primarily from the $8^3$ factor in the Wigner function for tetraquark system of Eq.~\ref{equ:mesoncc} relative to the single factor of 8 for two-body system of Eq.~\ref{equ:mesona}.

As mentioned in the introduction, the relative yields are more trustworthy from our coalescence calculation than the absolute yields themselves which would require more rigorous treatments of various aspects of particle production. Table~\ref{tab:ratio} lists the calculated ratios of $f_0(980)$ yield over $\pi^+$ yield with the various $f_0(980)$ configurations.

\begin{table}
\begin{center}
\caption{ \label{tab:ratio}Ratios of $f_0(980)$ yield over $\pi^+$ yield for different
configuration assumptions of the $f_0(980)$ in MB pp and pPb collisions at $\snn=5.0$~TeV.
The $\pi ^+$ yield from our calculations are $4.0539(6)$ and $6.6947(8)$
in MB pp and pPb collisions,respectively. 
}
\begin{tabular}{c c c c}
\hline
\rule{0pt}{10pt} $f_0(980)$ configuration \hspace{0.5cm} & \hspace{0.5cm} MB pp ($|y|<0.5$) \hspace{0.5cm} & \hspace{0.5cm} MB pPb ($0<y<0.5$) \hspace{0.5cm} \\
\hline
\rule{0pt}{10pt}
$u\bar{u}$ (p-wave) & $0.6558(2)$ & $0.4342(1)$\\
$s\bar{s}$ & $0.10632(5)$  & $0.0481(3)$ \\
$s\bar{s}u\bar{u}$ & $0.05412(3) $ & $0.0192(2)$\\
$K\bar{K}$ & $ 0.01698(2)$ & $0.01016(1)$\\
\hline 
\end{tabular}
\end{center}
\end{table}

\begin{figure}
\includegraphics[height=6cm]{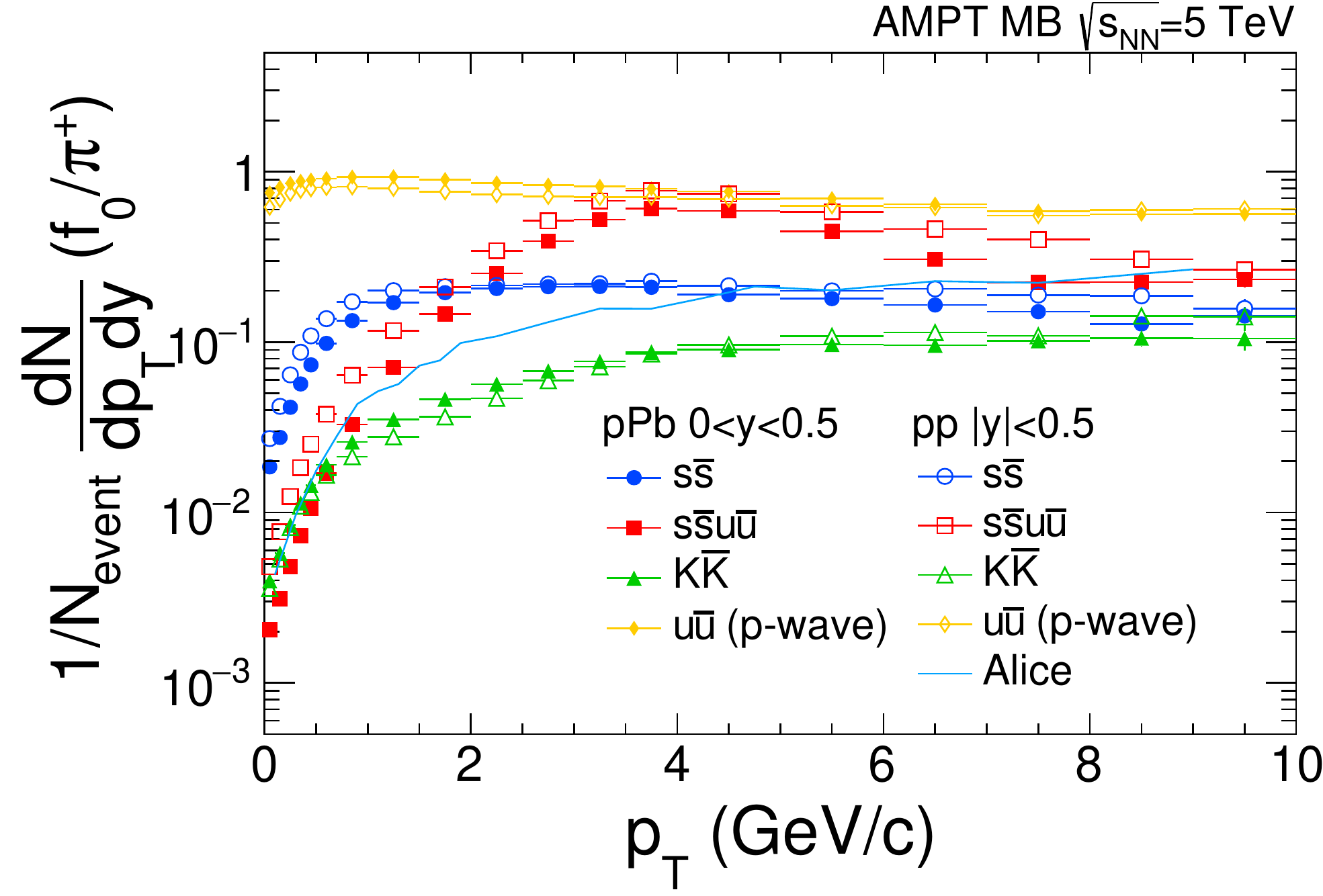}
\caption{\label{fig:pt_ratio_pion} The $p_T$ spectra ratio of $f_0(980)$ over $\pi^{+}$ ($u\bar{d}$) with different configuration assumptions for the $f_0(9880)$ in pp and pPb collisions. The input (anti-)quark phase-space information is taken from the corresponding AMPT simulations.}
\end{figure}

To study the spectral shape quantitatively, we show the ratio of the $p_T$ spectrum of the $f_0(980)$ of each configuration over that of the $\pi^+$s also calculated from our coalescence. 
This is shown in Fig.~\ref{fig:pt_ratio_pion}.
The ratio of $f_0(980)$ over AMPT $\pi^+$ differs significantly among the configurations but similar for pp and pPb collisions. 
We postulate that this self-normalized ratio would be a reliable prediction of the $f_0(980)$ to $\pi^+$ ratio in reality, and the shape of the ratio as a function of $p_T$ could be used to probe the $f_0(980)$ configuration.

The ALICE experiment has recently measured the $p_T$ spectrum of the $f_0(980)$ in inelastic pp collisions with center-of-mass energy $\sqrt{s}=5.02$~TeV at the LHC~\cite{ALICE:2022qnb}. 
Figure~\ref{fig:pt_ratio_pion} also shows the $f_0(980)/\pi$ ratio from ALICE measurements. 
It is not clear from the face value of the comparison what to conclude about the $f_0(980)$ quark content. However, given the simplicity of our coalescence model and the many physics questions regarding particle production in pp collisions, it will require significantly more detailed studies in order to draw firm conclusion. 

\begin{figure}
\subfloat[]{\includegraphics[width=0.5 \linewidth]{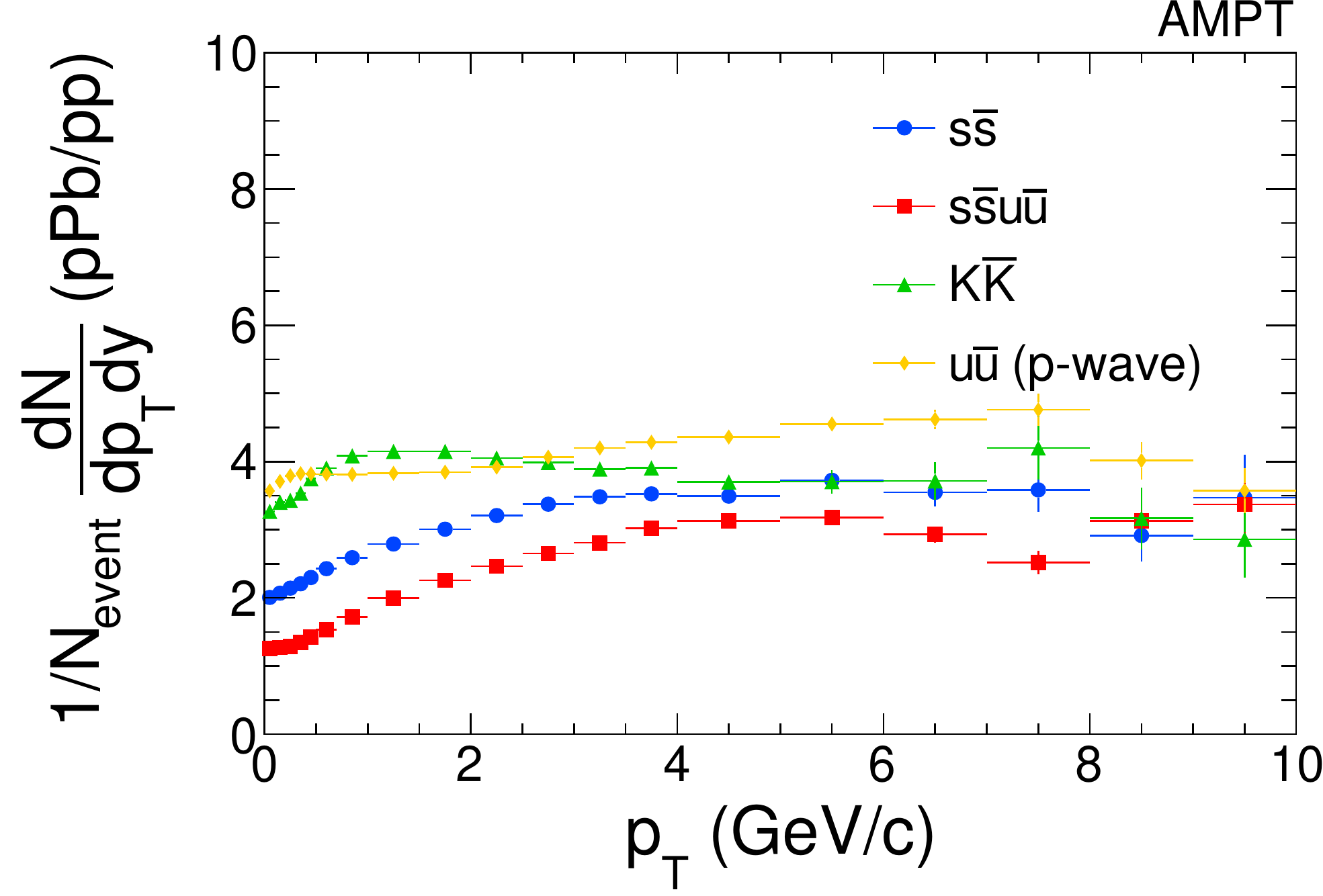}}
\subfloat[]{\includegraphics[width=0.5 \linewidth]{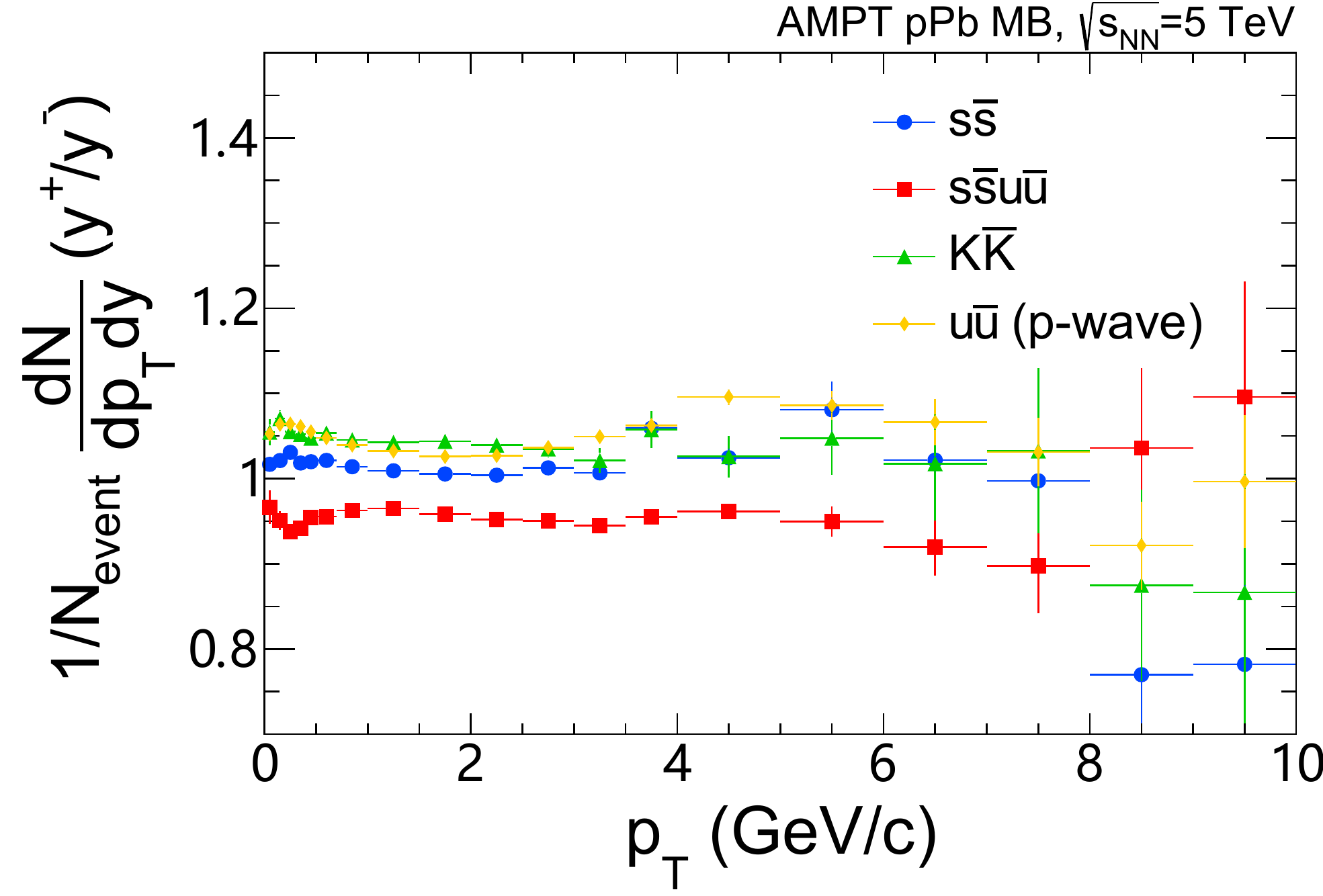}}
\caption{\label{fig:pt_ratio_etas_ctags} The $p_T$ specta ratio of $f_0(980)$ in Pb going side ($0<y<0.5$) over that in proton going side $-0.5<y<0$ for pPb collisions (left) and 
 over that in pp collisions ($|y|<0.5$) (right) with different configuration assumptions for the $f_0(980)$. The input (anti-)quark phase-space information is taken from the corresponding AMPT simulations.}
\end{figure}

Figure \ref{fig:pt_ratio_etas_ctags} left panel shows the ratio of $p_T$ spectrum of pPb collision over pp collision for each of the configurations. 
The yield from pPb collisions is about three times that of pp collisions, which is reasonable since the average parton number in pPb MB events is about three times that in pp MB events. 
The yield of hadrons is not proportional to the square of the number of partons, but rather linearly. 
This is because for each parton, it can only recombine with other partons within a certain distance. 
The ratio for $u\bar{u}$ is relatively flat. 
The ratio for $K\bar{K}$ increases at low $p_T$ and appears to flatten at high $p_T$. The ratios for $u\bar{u}$ and $K\bar{K}$ seem similar. 
The ratios of $s\bar{s}$ and $u\bar{u}s\bar{s}$ are distinct from those of $u\bar{u}$ and $K\bar{K}$. 
Both the ratios continuously increase with $p_T$. 
The ratio of $u\bar{u}s\bar{s}$ is lower than all other three. 
It appears that the pPb/pp ratio can potentially discriminate the $f_0(980)$ configuration.

The features of the pPb/pp ratio come from the parton densities of pp and pPb collisions that are significantly different. 
Since the parton densities also differ between the forward and backward rapidities in pPb collisions, it is attempting to examine the ratio of the backward/forward $p_T$ spectra in pPb collisions alone. 
This is shown in Fig. \ref{fig:pt_ratio_etas_ctags}(b). 
There are noticeable differences among the configurations; in particular, the ratio for tetraquark is lower than unity. 
This arises possibly from a larger phase-space volume on the Pb-going side, reducing the probability of tetraquark formation, than on the proton-going side.

\section{Conclusion}
We used a simple coalescence model with Gaussian Wigner function to generate $f_{0}(980)$ particles of four different configurations (${s\bar{s}}$, ${u\bar{u}s\bar{s}}$, $K\bar{K}$, $u\bar{u}$ p-wave), taking as input the phase space information of quarks (and Kaons) from the AMPT model. 
The yield and $p_T$ spectra are found to differ significantly among the configurations. 
This may be exploited to probe the $f_0(980)$ configuration by measuring $f_0(980)$ and other hadrons (such as pions) in a single collision system.
The ratios of the $p_T$ spectrum in pPb to that in pp collisions and those between forward and backward rapidities in pPb collisions are also studied for the various configurations and found to provide another means to probe the configuration of $f_0(980)$. 
These means, together with the number-of-constituent scaling of elliptic flow of the $f_0(980)$ \cite{Gu:2019oyz}, present promising prospects in determining the quark configuration of the $f_0(980)$.

As mentioned in the introduction, the validity of constituent quark coalescence in small systems of pp and pPb collisions is not without question.
Moreover, coalescence is not expected to apply at high $p_T$ where parton fragmentation is more relevant. 
Our calculated $p_T$ spectra of the $f_0(980)$ rely quantitatively on the phase-space distributions of partons from AMPT. 
Although the ratios of the $p_T$ spectra are presented as means to probe the $f_0(980)$ configuration, which are more robust than the spectra themselves, some model dependence can still exist. 
Such model dependence should be studied in the future. 
Our results depend also on the coalescence model parameters and how the coalescence is implemented (e.g. our simple single particle coalescence versus that in AMPT), and we leave such a study to future works. 

\section*{Acknowledgement}
FW thanks Dr.~Su Houng Lee for many fruitful discussions and for pointing out the possibility of $u\bar{u}$ p-wave configuration of the $f_0(980)$.
We thank Dr.~Zi-wei Lin for fruitful discussions.
This work was supported in part by the U.S.~Department of Energy (Grant No.~de-sc0012910).

\bibliography{apssamp}

\begin{thebibliography}{38}%
\makeatletter
\providecommand \@ifxundefined [1]{%
 \@ifx{#1\undefined}
}%
\providecommand \@ifnum [1]{%
 \ifnum #1\expandafter \@firstoftwo
 \else \expandafter \@secondoftwo
 \fi
}%
\providecommand \@ifx [1]{%
 \ifx #1\expandafter \@firstoftwo
 \else \expandafter \@secondoftwo
 \fi
}%
\providecommand \natexlab [1]{#1}%
\providecommand \enquote  [1]{``#1''}%
\providecommand \bibnamefont  [1]{#1}%
\providecommand \bibfnamefont [1]{#1}%
\providecommand \citenamefont [1]{#1}%
\providecommand \href@noop [0]{\@secondoftwo}%
\providecommand \href [0]{\begingroup \@sanitize@url \@href}%
\providecommand \@href[1]{\@@startlink{#1}\@@href}%
\providecommand \@@href[1]{\endgroup#1\@@endlink}%
\providecommand \@sanitize@url [0]{\catcode `\\12\catcode `\$12\catcode
  `\&12\catcode `\#12\catcode `\^12\catcode `\_12\catcode `\%12\relax}%
\providecommand \@@startlink[1]{}%
\providecommand \@@endlink[0]{}%
\providecommand \url  [0]{\begingroup\@sanitize@url \@url }%
\providecommand \@url [1]{\endgroup\@href {#1}{\urlprefix }}%
\providecommand \urlprefix  [0]{URL }%
\providecommand \Eprint [0]{\href }%
\providecommand \doibase [0]{http://dx.doi.org/}%
\providecommand \selectlanguage [0]{\@gobble}%
\providecommand \bibinfo  [0]{\@secondoftwo}%
\providecommand \bibfield  [0]{\@secondoftwo}%
\providecommand \translation [1]{[#1]}%
\providecommand \BibitemOpen [0]{}%
\providecommand \bibitemStop [0]{}%
\providecommand \bibitemNoStop [0]{.\EOS\space}%
\providecommand \EOS [0]{\spacefactor3000\relax}%
\providecommand \BibitemShut  [1]{\csname bibitem#1\endcsname}%
\let\auto@bib@innerbib\@empty
\bibitem [{\citenamefont {Jaffe}(2005)}]{Jaffe:2004ph}%
  \BibitemOpen
  \bibfield  {author} {\bibinfo {author} {\bibfnamefont {R.~L.}\ \bibnamefont
  {Jaffe}},\ }\bibfield  {booktitle} {\emph {\bibinfo {booktitle}
  {{Proceedings, 6th International Conference on Hyperons, charm and beauty
  hadrons (BEACH 2004): Chicago, USA, June 27-July 3, 2004}}},\ }\href
  {\doibase 10.1016/j.physrep.2004.11.005} {\bibfield  {journal} {\bibinfo
  {journal} {Phys. Rept.}\ }\textbf {\bibinfo {volume} {409}},\ \bibinfo
  {pages} {1} (\bibinfo {year} {2005})},\ \bibinfo {note} {[,191(2004)]},\
  \Eprint {http://arxiv.org/abs/hep-ph/0409065} {arXiv:hep-ph/0409065 [hep-ph]}
  \BibitemShut {NoStop}%
\bibitem [{\citenamefont {Protopopescu}\ \emph {et~al.}(1973)\citenamefont
  {Protopopescu}, \citenamefont {Alston-Garnjost}, \citenamefont
  {Barbaro-Galtieri}, \citenamefont {Flatte}, \citenamefont {Friedman},
  \citenamefont {Lasinski}, \citenamefont {Lynch}, \citenamefont {Rabin},\ and\
  \citenamefont {Solmitz}}]{Protopopescu:1973sh}%
  \BibitemOpen
  \bibfield  {author} {\bibinfo {author} {\bibfnamefont {S.~D.}\ \bibnamefont
  {Protopopescu}}, \bibinfo {author} {\bibfnamefont {M.}~\bibnamefont
  {Alston-Garnjost}}, \bibinfo {author} {\bibfnamefont {A.}~\bibnamefont
  {Barbaro-Galtieri}}, \bibinfo {author} {\bibfnamefont {S.~M.}\ \bibnamefont
  {Flatte}}, \bibinfo {author} {\bibfnamefont {J.~H.}\ \bibnamefont
  {Friedman}}, \bibinfo {author} {\bibfnamefont {T.~A.}\ \bibnamefont
  {Lasinski}}, \bibinfo {author} {\bibfnamefont {G.~R.}\ \bibnamefont {Lynch}},
  \bibinfo {author} {\bibfnamefont {M.~S.}\ \bibnamefont {Rabin}}, \ and\
  \bibinfo {author} {\bibfnamefont {F.~T.}\ \bibnamefont {Solmitz}},\ }\href
  {\doibase 10.1103/PhysRevD.7.1279} {\bibfield  {journal} {\bibinfo  {journal}
  {Phys. Rev.}\ }\textbf {\bibinfo {volume} {D7}},\ \bibinfo {pages} {1279}
  (\bibinfo {year} {1973})}\BibitemShut {NoStop}%
\bibitem [{\citenamefont {Hyams}\ \emph {et~al.}(1973)\citenamefont {Hyams}
  \emph {et~al.}}]{Hyams:1973zf}%
  \BibitemOpen
  \bibfield  {author} {\bibinfo {author} {\bibfnamefont {B.}~\bibnamefont
  {Hyams}} \emph {et~al.},\ }\href {\doibase 10.1016/0550-3213(73)90618-4}
  {\bibfield  {journal} {\bibinfo  {journal} {Nucl. Phys.}\ }\textbf {\bibinfo
  {volume} {B64}},\ \bibinfo {pages} {134} (\bibinfo {year}
  {1973})}\BibitemShut {NoStop}%
\bibitem [{\citenamefont {Grayer}\ \emph {et~al.}(1974)\citenamefont {Grayer}
  \emph {et~al.}}]{Grayer:1974cr}%
  \BibitemOpen
  \bibfield  {author} {\bibinfo {author} {\bibfnamefont {G.}~\bibnamefont
  {Grayer}} \emph {et~al.},\ }\href {\doibase 10.1016/0550-3213(74)90545-8}
  {\bibfield  {journal} {\bibinfo  {journal} {Nucl. Phys.}\ }\textbf {\bibinfo
  {volume} {B75}},\ \bibinfo {pages} {189} (\bibinfo {year}
  {1974})}\BibitemShut {NoStop}%
\bibitem [{\citenamefont {Bugg}(2004)}]{Bugg:2004xu}%
  \BibitemOpen
  \bibfield  {author} {\bibinfo {author} {\bibfnamefont {D.~V.}\ \bibnamefont
  {Bugg}},\ }\href {\doibase 10.1016/j.physrep.2004.03.008} {\bibfield
  {journal} {\bibinfo  {journal} {Phys. Rept.}\ }\textbf {\bibinfo {volume}
  {397}},\ \bibinfo {pages} {257} (\bibinfo {year} {2004})},\ \Eprint
  {http://arxiv.org/abs/hep-ex/0412045} {arXiv:hep-ex/0412045 [hep-ex]}
  \BibitemShut {NoStop}%
\bibitem [{\citenamefont {Klempt}\ and\ \citenamefont
  {Zaitsev}(2007)}]{Klempt:2007cp}%
  \BibitemOpen
  \bibfield  {author} {\bibinfo {author} {\bibfnamefont {E.}~\bibnamefont
  {Klempt}}\ and\ \bibinfo {author} {\bibfnamefont {A.}~\bibnamefont
  {Zaitsev}},\ }\href {\doibase 10.1016/j.physrep.2007.07.006} {\bibfield
  {journal} {\bibinfo  {journal} {Phys. Rept.}\ }\textbf {\bibinfo {volume}
  {454}},\ \bibinfo {pages} {1} (\bibinfo {year} {2007})},\ \Eprint
  {http://arxiv.org/abs/0708.4016} {arXiv:0708.4016 [hep-ph]} \BibitemShut
  {NoStop}%
\bibitem [{\citenamefont {Pelaez}(2016)}]{Pelaez:2015qba}%
  \BibitemOpen
  \bibfield  {author} {\bibinfo {author} {\bibfnamefont {J.~R.}\ \bibnamefont
  {Pelaez}},\ }\href {\doibase 10.1016/j.physrep.2016.09.001} {\bibfield
  {journal} {\bibinfo  {journal} {Phys. Rept.}\ }\textbf {\bibinfo {volume}
  {658}},\ \bibinfo {pages} {1} (\bibinfo {year} {2016})},\ \Eprint
  {http://arxiv.org/abs/1510.00653} {arXiv:1510.00653 [hep-ph]} \BibitemShut
  {NoStop}%
\bibitem [{\citenamefont {Adcox}\ \emph {et~al.}(2005)\citenamefont {Adcox}
  \emph {et~al.}}]{Adcox:2004mh}%
  \BibitemOpen
  \bibfield  {author} {\bibinfo {author} {\bibfnamefont {K.}~\bibnamefont
  {Adcox}} \emph {et~al.} (\bibinfo {collaboration} {PHENIX}),\ }\href
  {\doibase 10.1016/j.nuclphysa.2005.03.086} {\bibfield  {journal} {\bibinfo
  {journal} {Nucl. Phys.}\ }\textbf {\bibinfo {volume} {A757}},\ \bibinfo
  {pages} {184} (\bibinfo {year} {2005})},\ \Eprint
  {http://arxiv.org/abs/nucl-ex/0410003} {arXiv:nucl-ex/0410003 [nucl-ex]}
  \BibitemShut {NoStop}%
\bibitem [{\citenamefont {Arsene}\ \emph {et~al.}(2005)\citenamefont {Arsene}
  \emph {et~al.}}]{Arsene:2004fa}%
  \BibitemOpen
  \bibfield  {author} {\bibinfo {author} {\bibfnamefont {I.}~\bibnamefont
  {Arsene}} \emph {et~al.} (\bibinfo {collaboration} {BRAHMS}),\ }\href
  {\doibase 10.1016/j.nuclphysa.2005.02.130} {\bibfield  {journal} {\bibinfo
  {journal} {Nucl. Phys.}\ }\textbf {\bibinfo {volume} {A757}},\ \bibinfo
  {pages} {1} (\bibinfo {year} {2005})},\ \Eprint
  {http://arxiv.org/abs/nucl-ex/0410020} {arXiv:nucl-ex/0410020 [nucl-ex]}
  \BibitemShut {NoStop}%
\bibitem [{\citenamefont {Back}\ \emph {et~al.}(2005)\citenamefont {Back} \emph
  {et~al.}}]{Back:2004je}%
  \BibitemOpen
  \bibfield  {author} {\bibinfo {author} {\bibfnamefont {B.~B.}\ \bibnamefont
  {Back}} \emph {et~al.},\ }\href {\doibase 10.1016/j.nuclphysa.2005.03.084}
  {\bibfield  {journal} {\bibinfo  {journal} {Nucl. Phys.}\ }\textbf {\bibinfo
  {volume} {A757}},\ \bibinfo {pages} {28} (\bibinfo {year} {2005})},\ \Eprint
  {http://arxiv.org/abs/nucl-ex/0410022} {arXiv:nucl-ex/0410022 [nucl-ex]}
  \BibitemShut {NoStop}%
\bibitem [{\citenamefont {Adams}\ \emph {et~al.}(2005)\citenamefont {Adams}
  \emph {et~al.}}]{Adams:2005dq}%
  \BibitemOpen
  \bibfield  {author} {\bibinfo {author} {\bibfnamefont {J.}~\bibnamefont
  {Adams}} \emph {et~al.} (\bibinfo {collaboration} {STAR}),\ }\href {\doibase
  10.1016/j.nuclphysa.2005.03.085} {\bibfield  {journal} {\bibinfo  {journal}
  {Nucl. Phys.}\ }\textbf {\bibinfo {volume} {A757}},\ \bibinfo {pages} {102}
  (\bibinfo {year} {2005})},\ \Eprint {http://arxiv.org/abs/nucl-ex/0501009}
  {arXiv:nucl-ex/0501009 [nucl-ex]} \BibitemShut {NoStop}%
\bibitem [{\citenamefont {Muller}\ \emph {et~al.}(2012)\citenamefont {Muller},
  \citenamefont {Schukraft},\ and\ \citenamefont {Wyslouch}}]{Muller:2012zq}%
  \BibitemOpen
  \bibfield  {author} {\bibinfo {author} {\bibfnamefont {B.}~\bibnamefont
  {Muller}}, \bibinfo {author} {\bibfnamefont {J.}~\bibnamefont {Schukraft}}, \
  and\ \bibinfo {author} {\bibfnamefont {B.}~\bibnamefont {Wyslouch}},\ }\href
  {\doibase 10.1146/annurev-nucl-102711-094910} {\bibfield  {journal} {\bibinfo
   {journal} {Ann. Rev. Nucl. Part. Sci.}\ }\textbf {\bibinfo {volume} {62}},\
  \bibinfo {pages} {361} (\bibinfo {year} {2012})},\ \Eprint
  {http://arxiv.org/abs/1202.3233} {arXiv:1202.3233 [hep-ex]} \BibitemShut
  {NoStop}%
\bibitem [{\citenamefont {Dover}\ \emph {et~al.}(1991)\citenamefont {Dover},
  \citenamefont {Heinz}, \citenamefont {Schnedermann},\ and\ \citenamefont
  {Zimanyi}}]{Dover:1991zn}%
  \BibitemOpen
  \bibfield  {author} {\bibinfo {author} {\bibfnamefont {C.~B.}\ \bibnamefont
  {Dover}}, \bibinfo {author} {\bibfnamefont {U.~W.}\ \bibnamefont {Heinz}},
  \bibinfo {author} {\bibfnamefont {E.}~\bibnamefont {Schnedermann}}, \ and\
  \bibinfo {author} {\bibfnamefont {J.}~\bibnamefont {Zimanyi}},\ }\href
  {\doibase 10.1103/PhysRevC.44.1636} {\bibfield  {journal} {\bibinfo
  {journal} {Phys. Rev.}\ }\textbf {\bibinfo {volume} {C44}},\ \bibinfo {pages}
  {1636} (\bibinfo {year} {1991})}\BibitemShut {NoStop}%
\bibitem [{\citenamefont {Fries}\ \emph {et~al.}(2008)\citenamefont {Fries},
  \citenamefont {Greco},\ and\ \citenamefont {Sorensen}}]{Fries:2008hs}%
  \BibitemOpen
  \bibfield  {author} {\bibinfo {author} {\bibfnamefont {R.~J.}\ \bibnamefont
  {Fries}}, \bibinfo {author} {\bibfnamefont {V.}~\bibnamefont {Greco}}, \ and\
  \bibinfo {author} {\bibfnamefont {P.}~\bibnamefont {Sorensen}},\ }\href
  {\doibase 10.1146/annurev.nucl.58.110707.171134} {\bibfield  {journal}
  {\bibinfo  {journal} {Ann.Rev.Nucl.Part.Sci.}\ }\textbf {\bibinfo {volume}
  {58}},\ \bibinfo {pages} {177} (\bibinfo {year} {2008})},\ \Eprint
  {http://arxiv.org/abs/0807.4939} {arXiv:0807.4939 [nucl-th]} \BibitemShut
  {NoStop}%
\bibitem [{\citenamefont {Butler}\ and\ \citenamefont
  {Pearson}(1963)}]{Butler:1963pp}%
  \BibitemOpen
  \bibfield  {author} {\bibinfo {author} {\bibfnamefont {S.~T.}\ \bibnamefont
  {Butler}}\ and\ \bibinfo {author} {\bibfnamefont {C.~A.}\ \bibnamefont
  {Pearson}},\ }\href {\doibase 10.1103/PhysRev.129.836} {\bibfield  {journal}
  {\bibinfo  {journal} {Phys. Rev.}\ }\textbf {\bibinfo {volume} {129}},\
  \bibinfo {pages} {836} (\bibinfo {year} {1963})}\BibitemShut {NoStop}%
\bibitem [{\citenamefont {Hwa}\ and\ \citenamefont {Yang}(2003)}]{Hwa:2002tu}%
  \BibitemOpen
  \bibfield  {author} {\bibinfo {author} {\bibfnamefont {R.~C.}\ \bibnamefont
  {Hwa}}\ and\ \bibinfo {author} {\bibfnamefont {C.~B.}\ \bibnamefont {Yang}},\
  }\href {\doibase 10.1103/PhysRevC.67.034902} {\bibfield  {journal} {\bibinfo
  {journal} {Phys. Rev.}\ }\textbf {\bibinfo {volume} {C67}},\ \bibinfo {pages}
  {034902} (\bibinfo {year} {2003})},\ \Eprint
  {http://arxiv.org/abs/nucl-th/0211010} {arXiv:nucl-th/0211010 [nucl-th]}
  \BibitemShut {NoStop}%
\bibitem [{\citenamefont {Fries}\ \emph
  {et~al.}(2003{\natexlab{a}})\citenamefont {Fries}, \citenamefont {Muller},
  \citenamefont {Nonaka},\ and\ \citenamefont {Bass}}]{Fries:2003vb}%
  \BibitemOpen
  \bibfield  {author} {\bibinfo {author} {\bibfnamefont {R.~J.}\ \bibnamefont
  {Fries}}, \bibinfo {author} {\bibfnamefont {B.}~\bibnamefont {Muller}},
  \bibinfo {author} {\bibfnamefont {C.}~\bibnamefont {Nonaka}}, \ and\ \bibinfo
  {author} {\bibfnamefont {S.~A.}\ \bibnamefont {Bass}},\ }\href {\doibase
  10.1103/PhysRevLett.90.202303} {\bibfield  {journal} {\bibinfo  {journal}
  {Phys. Rev. Lett.}\ }\textbf {\bibinfo {volume} {90}},\ \bibinfo {pages}
  {202303} (\bibinfo {year} {2003}{\natexlab{a}})},\ \Eprint
  {http://arxiv.org/abs/nucl-th/0301087} {arXiv:nucl-th/0301087 [nucl-th]}
  \BibitemShut {NoStop}%
\bibitem [{\citenamefont {Fries}\ \emph
  {et~al.}(2003{\natexlab{b}})\citenamefont {Fries}, \citenamefont {Muller},
  \citenamefont {Nonaka},\ and\ \citenamefont {Bass}}]{Fries:2003kq}%
  \BibitemOpen
  \bibfield  {author} {\bibinfo {author} {\bibfnamefont {R.~J.}\ \bibnamefont
  {Fries}}, \bibinfo {author} {\bibfnamefont {B.}~\bibnamefont {Muller}},
  \bibinfo {author} {\bibfnamefont {C.}~\bibnamefont {Nonaka}}, \ and\ \bibinfo
  {author} {\bibfnamefont {S.~A.}\ \bibnamefont {Bass}},\ }\href {\doibase
  10.1103/PhysRevC.68.044902} {\bibfield  {journal} {\bibinfo  {journal} {Phys.
  Rev.}\ }\textbf {\bibinfo {volume} {C68}},\ \bibinfo {pages} {044902}
  (\bibinfo {year} {2003}{\natexlab{b}})},\ \Eprint
  {http://arxiv.org/abs/nucl-th/0306027} {arXiv:nucl-th/0306027 [nucl-th]}
  \BibitemShut {NoStop}%
\bibitem [{\citenamefont {Greco}\ \emph
  {et~al.}(2003{\natexlab{a}})\citenamefont {Greco}, \citenamefont {Ko},\ and\
  \citenamefont {Levai}}]{Greco:2003xt}%
  \BibitemOpen
  \bibfield  {author} {\bibinfo {author} {\bibfnamefont {V.}~\bibnamefont
  {Greco}}, \bibinfo {author} {\bibfnamefont {C.~M.}\ \bibnamefont {Ko}}, \
  and\ \bibinfo {author} {\bibfnamefont {P.}~\bibnamefont {Levai}},\ }\href
  {\doibase 10.1103/PhysRevLett.90.202302} {\bibfield  {journal} {\bibinfo
  {journal} {Phys. Rev. Lett.}\ }\textbf {\bibinfo {volume} {90}},\ \bibinfo
  {pages} {202302} (\bibinfo {year} {2003}{\natexlab{a}})},\ \Eprint
  {http://arxiv.org/abs/nucl-th/0301093} {arXiv:nucl-th/0301093 [nucl-th]}
  \BibitemShut {NoStop}%
\bibitem [{\citenamefont {Greco}\ \emph
  {et~al.}(2003{\natexlab{b}})\citenamefont {Greco}, \citenamefont {Ko},\ and\
  \citenamefont {Levai}}]{Greco:2003mm}%
  \BibitemOpen
  \bibfield  {author} {\bibinfo {author} {\bibfnamefont {V.}~\bibnamefont
  {Greco}}, \bibinfo {author} {\bibfnamefont {C.~M.}\ \bibnamefont {Ko}}, \
  and\ \bibinfo {author} {\bibfnamefont {P.}~\bibnamefont {Levai}},\ }\href
  {\doibase 10.1103/PhysRevC.68.034904} {\bibfield  {journal} {\bibinfo
  {journal} {Phys. Rev.}\ }\textbf {\bibinfo {volume} {C68}},\ \bibinfo {pages}
  {034904} (\bibinfo {year} {2003}{\natexlab{b}})},\ \Eprint
  {http://arxiv.org/abs/nucl-th/0305024} {arXiv:nucl-th/0305024 [nucl-th]}
  \BibitemShut {NoStop}%
\bibitem [{\citenamefont {Hwa}\ and\ \citenamefont {Yang}(2004)}]{Hwa:2004ng}%
  \BibitemOpen
  \bibfield  {author} {\bibinfo {author} {\bibfnamefont {R.~C.}\ \bibnamefont
  {Hwa}}\ and\ \bibinfo {author} {\bibfnamefont {C.~B.}\ \bibnamefont {Yang}},\
  }\href {\doibase 10.1103/PhysRevC.70.024905} {\bibfield  {journal} {\bibinfo
  {journal} {Phys. Rev.}\ }\textbf {\bibinfo {volume} {C70}},\ \bibinfo {pages}
  {024905} (\bibinfo {year} {2004})},\ \Eprint
  {http://arxiv.org/abs/nucl-th/0401001} {arXiv:nucl-th/0401001 [nucl-th]}
  \BibitemShut {NoStop}%
\bibitem [{\citenamefont {Minissale}\ \emph {et~al.}(2015)\citenamefont
  {Minissale}, \citenamefont {Scardina},\ and\ \citenamefont
  {Greco}}]{Minissale:2015zwa}%
  \BibitemOpen
  \bibfield  {author} {\bibinfo {author} {\bibfnamefont {V.}~\bibnamefont
  {Minissale}}, \bibinfo {author} {\bibfnamefont {F.}~\bibnamefont {Scardina}},
  \ and\ \bibinfo {author} {\bibfnamefont {V.}~\bibnamefont {Greco}},\ }\href
  {\doibase 10.1103/PhysRevC.92.054904} {\bibfield  {journal} {\bibinfo
  {journal} {Phys. Rev.}\ }\textbf {\bibinfo {volume} {C92}},\ \bibinfo {pages}
  {054904} (\bibinfo {year} {2015})},\ \Eprint
  {http://arxiv.org/abs/1502.06213} {arXiv:1502.06213 [nucl-th]} \BibitemShut
  {NoStop}%
\bibitem [{\citenamefont {Molnar}\ and\ \citenamefont
  {Voloshin}(2003)}]{Molnar:2003ff}%
  \BibitemOpen
  \bibfield  {author} {\bibinfo {author} {\bibfnamefont {D.}~\bibnamefont
  {Molnar}}\ and\ \bibinfo {author} {\bibfnamefont {S.~A.}\ \bibnamefont
  {Voloshin}},\ }\href {\doibase 10.1103/PhysRevLett.91.092301} {\bibfield
  {journal} {\bibinfo  {journal} {Phys.Rev.Lett.}\ }\textbf {\bibinfo {volume}
  {91}},\ \bibinfo {pages} {092301} (\bibinfo {year} {2003})},\ \Eprint
  {http://arxiv.org/abs/nucl-th/0302014} {arXiv:nucl-th/0302014 [nucl-th]}
  \BibitemShut {NoStop}%
\bibitem [{\citenamefont {Gu}\ \emph {et~al.}(2020)\citenamefont {Gu},
  \citenamefont {Edmonds}, \citenamefont {Zhao},\ and\ \citenamefont
  {Wang}}]{Gu:2019oyz}%
  \BibitemOpen
  \bibfield  {author} {\bibinfo {author} {\bibfnamefont {A.}~\bibnamefont
  {Gu}}, \bibinfo {author} {\bibfnamefont {T.}~\bibnamefont {Edmonds}},
  \bibinfo {author} {\bibfnamefont {J.}~\bibnamefont {Zhao}}, \ and\ \bibinfo
  {author} {\bibfnamefont {F.}~\bibnamefont {Wang}},\ }\href {\doibase
  10.1103/PhysRevC.101.024908} {\bibfield  {journal} {\bibinfo  {journal}
  {Phys. Rev. C}\ }\textbf {\bibinfo {volume} {101}},\ \bibinfo {pages}
  {024908} (\bibinfo {year} {2020})},\ \Eprint
  {http://arxiv.org/abs/1902.07152} {arXiv:1902.07152 [nucl-ex]} \BibitemShut
  {NoStop}%
\bibitem [{\citenamefont {Lin}\ \emph {et~al.}(2005)\citenamefont {Lin},
  \citenamefont {Ko}, \citenamefont {Li}, \citenamefont {Zhang},\ and\
  \citenamefont {Pal}}]{Lin:2004en}%
  \BibitemOpen
  \bibfield  {author} {\bibinfo {author} {\bibfnamefont {Z.-W.}\ \bibnamefont
  {Lin}}, \bibinfo {author} {\bibfnamefont {C.~M.}\ \bibnamefont {Ko}},
  \bibinfo {author} {\bibfnamefont {B.-A.}\ \bibnamefont {Li}}, \bibinfo
  {author} {\bibfnamefont {B.}~\bibnamefont {Zhang}}, \ and\ \bibinfo {author}
  {\bibfnamefont {S.}~\bibnamefont {Pal}},\ }\href {\doibase
  10.1103/PhysRevC.72.064901} {\bibfield  {journal} {\bibinfo  {journal} {Phys.
  Rev.}\ }\textbf {\bibinfo {volume} {C72}},\ \bibinfo {pages} {064901}
  (\bibinfo {year} {2005})},\ \Eprint {http://arxiv.org/abs/nucl-th/0411110}
  {arXiv:nucl-th/0411110 [nucl-th]} \BibitemShut {NoStop}%
\bibitem [{\citenamefont {Lin}\ and\ \citenamefont
  {Molnar}(2003)}]{Lin:2003jy}%
  \BibitemOpen
  \bibfield  {author} {\bibinfo {author} {\bibfnamefont {Z.-w.}\ \bibnamefont
  {Lin}}\ and\ \bibinfo {author} {\bibfnamefont {D.}~\bibnamefont {Molnar}},\
  }\href {\doibase 10.1103/PhysRevC.68.044901} {\bibfield  {journal} {\bibinfo
  {journal} {Phys. Rev.}\ }\textbf {\bibinfo {volume} {C68}},\ \bibinfo {pages}
  {044901} (\bibinfo {year} {2003})},\ \Eprint
  {http://arxiv.org/abs/nucl-th/0304045} {arXiv:nucl-th/0304045 [nucl-th]}
  \BibitemShut {NoStop}%
\bibitem [{\citenamefont {He}\ and\ \citenamefont {Lin}(2017)}]{He:2017tla}%
  \BibitemOpen
  \bibfield  {author} {\bibinfo {author} {\bibfnamefont {Y.}~\bibnamefont
  {He}}\ and\ \bibinfo {author} {\bibfnamefont {Z.-W.}\ \bibnamefont {Lin}},\
  }\href {\doibase 10.1103/PhysRevC.96.014910} {\bibfield  {journal} {\bibinfo
  {journal} {Phys. Rev.}\ }\textbf {\bibinfo {volume} {C96}},\ \bibinfo {pages}
  {014910} (\bibinfo {year} {2017})},\ \Eprint
  {http://arxiv.org/abs/1703.02673} {arXiv:1703.02673 [nucl-th]} \BibitemShut
  {NoStop}%
\bibitem [{\citenamefont {Cho}\ \emph {et~al.}(2011)\citenamefont {Cho} \emph
  {et~al.}}]{Cho:2010db}%
  \BibitemOpen
  \bibfield  {author} {\bibinfo {author} {\bibfnamefont {S.}~\bibnamefont
  {Cho}} \emph {et~al.} (\bibinfo {collaboration} {ExHIC}),\ }\href {\doibase
  10.1103/PhysRevLett.106.212001} {\bibfield  {journal} {\bibinfo  {journal}
  {Phys. Rev. Lett.}\ }\textbf {\bibinfo {volume} {106}},\ \bibinfo {pages}
  {212001} (\bibinfo {year} {2011})},\ \Eprint {http://arxiv.org/abs/1011.0852}
  {arXiv:1011.0852 [nucl-th]} \BibitemShut {NoStop}%
\bibitem [{\citenamefont {Zhang}\ \emph {et~al.}(2021)\citenamefont {Zhang},
  \citenamefont {Zheng}, \citenamefont {Shi},\ and\ \citenamefont
  {Lin}}]{Zhang:2021vvp}%
  \BibitemOpen
  \bibfield  {author} {\bibinfo {author} {\bibfnamefont {C.}~\bibnamefont
  {Zhang}}, \bibinfo {author} {\bibfnamefont {L.}~\bibnamefont {Zheng}},
  \bibinfo {author} {\bibfnamefont {S.}~\bibnamefont {Shi}}, \ and\ \bibinfo
  {author} {\bibfnamefont {Z.-W.}\ \bibnamefont {Lin}},\ }\href {\doibase
  10.1103/PhysRevC.104.014908} {\bibfield  {journal} {\bibinfo  {journal}
  {Phys. Rev. C}\ }\textbf {\bibinfo {volume} {104}},\ \bibinfo {pages}
  {014908} (\bibinfo {year} {2021})},\ \Eprint
  {http://arxiv.org/abs/2103.10815} {arXiv:2103.10815 [nucl-th]} \BibitemShut
  {NoStop}%
\bibitem [{\citenamefont {Zhang}\ \emph {et~al.}(2022)\citenamefont {Zhang},
  \citenamefont {Zheng}, \citenamefont {Shi},\ and\ \citenamefont
  {Lin}}]{Zhang:2022fum}%
  \BibitemOpen
  \bibfield  {author} {\bibinfo {author} {\bibfnamefont {C.}~\bibnamefont
  {Zhang}}, \bibinfo {author} {\bibfnamefont {L.}~\bibnamefont {Zheng}},
  \bibinfo {author} {\bibfnamefont {S.}~\bibnamefont {Shi}}, \ and\ \bibinfo
  {author} {\bibfnamefont {Z.-W.}\ \bibnamefont {Lin}},\ }\href@noop {} {\
  (\bibinfo {year} {2022})},\ \Eprint {http://arxiv.org/abs/2210.07767}
  {arXiv:2210.07767 [nucl-th]} \BibitemShut {NoStop}%
\bibitem [{\citenamefont {Han}\ \emph {et~al.}(2016)\citenamefont {Han},
  \citenamefont {Fries},\ and\ \citenamefont {Ko}}]{Han:2016uhh}%
  \BibitemOpen
  \bibfield  {author} {\bibinfo {author} {\bibfnamefont {K.~C.}\ \bibnamefont
  {Han}}, \bibinfo {author} {\bibfnamefont {R.~J.}\ \bibnamefont {Fries}}, \
  and\ \bibinfo {author} {\bibfnamefont {C.~M.}\ \bibnamefont {Ko}},\ }\href
  {\doibase 10.1103/PhysRevC.93.045207} {\bibfield  {journal} {\bibinfo
  {journal} {Phys. Rev.}\ }\textbf {\bibinfo {volume} {C93}},\ \bibinfo {pages}
  {045207} (\bibinfo {year} {2016})},\ \Eprint
  {http://arxiv.org/abs/1601.00708} {arXiv:1601.00708 [nucl-th]} \BibitemShut
  {NoStop}%
\bibitem [{\citenamefont {Ali}\ \emph {et~al.}(2017)\citenamefont {Ali},
  \citenamefont {Lange},\ and\ \citenamefont {Stone}}]{Ali:2017jda}%
  \BibitemOpen
  \bibfield  {author} {\bibinfo {author} {\bibfnamefont {A.}~\bibnamefont
  {Ali}}, \bibinfo {author} {\bibfnamefont {J.~S.}\ \bibnamefont {Lange}}, \
  and\ \bibinfo {author} {\bibfnamefont {S.}~\bibnamefont {Stone}},\ }\href
  {\doibase 10.1016/j.ppnp.2017.08.003} {\bibfield  {journal} {\bibinfo
  {journal} {Prog. Part. Nucl. Phys.}\ }\textbf {\bibinfo {volume} {97}},\
  \bibinfo {pages} {123} (\bibinfo {year} {2017})},\ \Eprint
  {http://arxiv.org/abs/1706.00610} {arXiv:1706.00610 [hep-ph]} \BibitemShut
  {NoStop}%
\bibitem [{\citenamefont {Chen}\ \emph {et~al.}(2016)\citenamefont {Chen},
  \citenamefont {Chen}, \citenamefont {Liu},\ and\ \citenamefont
  {Zhu}}]{Chen:2016qju}%
  \BibitemOpen
  \bibfield  {author} {\bibinfo {author} {\bibfnamefont {H.-X.}\ \bibnamefont
  {Chen}}, \bibinfo {author} {\bibfnamefont {W.}~\bibnamefont {Chen}}, \bibinfo
  {author} {\bibfnamefont {X.}~\bibnamefont {Liu}}, \ and\ \bibinfo {author}
  {\bibfnamefont {S.-L.}\ \bibnamefont {Zhu}},\ }\href {\doibase
  10.1016/j.physrep.2016.05.004} {\bibfield  {journal} {\bibinfo  {journal}
  {Phys. Rept.}\ }\textbf {\bibinfo {volume} {639}},\ \bibinfo {pages} {1}
  (\bibinfo {year} {2016})},\ \Eprint {http://arxiv.org/abs/1601.02092}
  {arXiv:1601.02092 [hep-ph]} \BibitemShut {NoStop}%
\bibitem [{\citenamefont {Brink}\ and\ \citenamefont
  {Stancu}(1998)}]{PhysRevD.57.6778}%
  \BibitemOpen
  \bibfield  {author} {\bibinfo {author} {\bibfnamefont {D.~M.}\ \bibnamefont
  {Brink}}\ and\ \bibinfo {author} {\bibfnamefont {F.}~\bibnamefont {Stancu}},\
  }\href {\doibase 10.1103/PhysRevD.57.6778} {\bibfield  {journal} {\bibinfo
  {journal} {Phys. Rev. D}\ }\textbf {\bibinfo {volume} {57}},\ \bibinfo
  {pages} {6778} (\bibinfo {year} {1998})}\BibitemShut {NoStop}%
\bibitem [{\citenamefont {Shi}\ \emph {et~al.}(2013)\citenamefont {Shi},
  \citenamefont {Guo},\ and\ \citenamefont {Zhuang}}]{Shi:2013rga}%
  \BibitemOpen
  \bibfield  {author} {\bibinfo {author} {\bibfnamefont {S.}~\bibnamefont
  {Shi}}, \bibinfo {author} {\bibfnamefont {X.}~\bibnamefont {Guo}}, \ and\
  \bibinfo {author} {\bibfnamefont {P.}~\bibnamefont {Zhuang}},\ }\href
  {\doibase 10.1103/PhysRevD.88.014021} {\bibfield  {journal} {\bibinfo
  {journal} {Phys. Rev.}\ }\textbf {\bibinfo {volume} {D88}},\ \bibinfo {pages}
  {014021} (\bibinfo {year} {2013})},\ \Eprint {http://arxiv.org/abs/1306.1896}
  {arXiv:1306.1896 [nucl-th]} \BibitemShut {NoStop}%
\bibitem [{\citenamefont {Mrowczynski}(1987)}]{Mrowczynski:1987oid}%
  \BibitemOpen
  \bibfield  {author} {\bibinfo {author} {\bibfnamefont {S.}~\bibnamefont
  {Mrowczynski}},\ }\href {\doibase 10.1088/0305-4616/13/9/011} {\bibfield
  {journal} {\bibinfo  {journal} {J. Phys.}\ }\textbf {\bibinfo {volume}
  {G13}},\ \bibinfo {pages} {1089} (\bibinfo {year} {1987})}\BibitemShut
  {NoStop}%
\bibitem [{\citenamefont {Acharya}\ \emph {et~al.}(2020)\citenamefont {Acharya}
  \emph {et~al.}}]{ALICE:2019hno}%
  \BibitemOpen
  \bibfield  {author} {\bibinfo {author} {\bibfnamefont {S.}~\bibnamefont
  {Acharya}} \emph {et~al.} (\bibinfo {collaboration} {ALICE}),\ }\href
  {\doibase 10.1103/PhysRevC.101.044907} {\bibfield  {journal} {\bibinfo
  {journal} {Phys. Rev. C}\ }\textbf {\bibinfo {volume} {101}},\ \bibinfo
  {pages} {044907} (\bibinfo {year} {2020})},\ \Eprint
  {http://arxiv.org/abs/1910.07678} {arXiv:1910.07678 [nucl-ex]} \BibitemShut
  {NoStop}%
\bibitem [{\citenamefont {Acharya}\ \emph {et~al.}(2023)\citenamefont {Acharya}
  \emph {et~al.}}]{ALICE:2022qnb}%
  \BibitemOpen
  \bibfield  {author} {\bibinfo {author} {\bibfnamefont {S.}~\bibnamefont
  {Acharya}} \emph {et~al.} (\bibinfo {collaboration} {ALICE}),\ }\href
  {\doibase 10.1016/j.physletb.2022.137644} {\bibfield  {journal} {\bibinfo
  {journal} {Phys. Lett. B}\ }\textbf {\bibinfo {volume} {846}},\ \bibinfo
  {pages} {137644} (\bibinfo {year} {2023})},\ \Eprint
  {http://arxiv.org/abs/2206.06216} {arXiv:2206.06216 [nucl-ex]} \BibitemShut
  {NoStop}%
\end{thebibliography}%
\end{document}